      [1999/12/01 v1.4c Il Nuovo Cimento]
\documentclass{cimento}

\usepackage{verbatim}
\usepackage{natbib}
\usepackage{amsmath}
\usepackage{amssymb}
\usepackage{amsbsy}
\usepackage{lscape}
\usepackage{graphicx}

\newcommand{\Lya}{\mbox{Ly$\alpha$}}
\newcommand{\Gbkg}{\mbox{$\Gamma^{bkg}$}}

\title{A Direct Precision Measurement of the Intergalactic Lyman-$\alpha$ Opacity at $2\leq z\leq4.2$\thanks{Some of the data presented herein were obtained at the W.M. Keck
Observatory, which is operated as a scientific partnership among the
California Institute of Technology, the University of California and the
National Aeronautics and Space Administration. The Observatory was made
possible by the generous financial support of the W.M. Keck Foundation.
Other data analyzed in this work were gathered with the 6.5 meter Magellan Telescopes located at Las Campanas Observatory, Chile.}}
\author{Claude-Andr\'e Faucher-Gigu\`ere\from{harvard_astro}\ETC,
Jason X. Prochaska\from{ucsc},
Adam Lidz\from{harvard_astro},
Lars Hernquist\from{harvard_astro} \atque
Matias Zaldarriaga\from{harvard_astro}}
\instlist{\inst{harvard_astro} Department of Astronomy, Harvard University, 60 Garden St, MS-10, Cambridge, MA, 02138, USA; cgiguere@cfa.harvard.edu.  
\inst{ucsc} Department of Astronomy and Astrophysics, UCO/Lick Observatory; University of California, 1156 High Street, Santa Cruz, CA, 95064, USA.}
\PACSes{\PACSit{98.62.Ra, 98.54.Aj, 98.80.Es}{Intergalactic opacity, \Lya~forest, quasars}}
\begin{document}

\maketitle

\begin{abstract}
We directly measure the evolution of the intergalactic Lyman-$\alpha$ effective optical depth, $\tau_{\rm eff}$, over the redshift range $2 \leq z \leq 4.2$ from a sample of 86 high-resolution, high-signal-to-noise quasar spectra obtained with Keck/ESI, Keck/HIRES, and Magellan/MIKE. 
We find that our estimates of the quasar continuum levels in the \Lya~forest obtained by spline fitting are systematically biased low, but that this bias can be accounted for using mock spectra.
The mean fractional error $\langle \Delta C/C_{\rm true} \rangle$ is $<1\%$ at $z=2$, 4\% at $z=3$, and 12\% at $z=4$.
We provide estimates of the level of absorption arising from metals in the \Lya~forest based on both direct and statistical metal removal results in the literature, finding that this contribution is $\approx6-9\%$ at $z=3$ and decreases monotonically with redshift.
The high precision of our measurement indicates significant departures from the best-fit power-law redshift evolution, particularly near $z=3.2$.
\end{abstract}

\section{INTRODUCTION}
The evolution of the intergalactic medium (IGM) as traced by the \Lya~forest provides a powerful record of the thermal and radiative history of the Universe.
This power owes to our ability to measure the \Lya~opacity of the
IGM as a function of redshift, as well as to the relatively simple physics of the \Lya~forest.
In fact, cosmological simulations in which the forest arises from absorption by smooth density fluctuations imposed on the warm photoionized IGM as a natural consequence of hierarchical structure formation within cold dark matter models \citep[e.g.,][]{1996ApJ...457L..51H}, have been remarkably successful at reproducing the properties of the absorption observed in actual quasar spectra.
This synergy between theory and observations make the \Lya~forest a particularly compelling probe of the diffuse Universe.\\ \\
In this work, we present a direct precision measurement of the effective \Lya~optical depth and its evolution over the redshift range $2\leq z \leq4.2$ from a sample of 86 high-resolution, high signal-to-noise quasar spectra obtained obtained with Keck/ESI (16), Keck/HIRES (44), and with Magellan/MIKE (26).
The full details have been reported in \citep{2007arXiv0709.2382F}.
We assume a cosmology with $(\Omega_{m},~\Omega_{b},~\Omega_{\Lambda},~h,~\sigma_{8})=(0.27,~0.046,~0.73,~0.7,~0.8)$ \citep[][]{2007ApJS..170..377S}.

\section{ANALYSIS}
\subsection{Method}
Let $F_{\rm abs}(\lambda)$ be a quasar's absolute flux and $C(\lambda)$ be its continuum (unabsorbed) level as a function of observed wavelength $\lambda$.
The corresponding redshift of 
\Lya~absorption is just $z_{{\rm Ly}\alpha}\equiv \lambda/\lambda_{\Lya}-1$, where $\lambda_{\Lya}=1216$ \AA.
Dropping the \Lya~subscript on $z$, we define the transmission as $F(z)\equiv F_{\rm abs}(z)/C(z)$.
Letting $\langle F \rangle$ be the ensemble average over lines of sight, the effective optical depth is defined as $\tau_{\rm eff}(z) \equiv -\ln{\left[\langle F \rangle(z)\right]}$
The measurement consists of the following steps:
\begin{enumerate}
\item Estimate the continuum level in the \Lya~forest of each quasar spectrum by fitting a cubic spline through its transmission peaks and calculate the corresponding $F$;
\item apply masks to the data: quasar proximity regions, higher-order Lyman series forests, SLLS, DLAs, and associated metal lines;
\item bin the remaining \Lya~forest pixels in redshift intervals;
\item for each redshift bin, average the pixels to estimate $\langle F \rangle$ and calculate $\tau_{\rm eff}$;
\item correct this ``raw'' measurement for redshift-dependent continuum estimation error and, optionally, for metal absorption.
\end{enumerate}

\subsection{Continuum Correction}
As the redshift increases, the IGM density increases and, as a result, the transmission peaks reaching unity become increasingly rare, causing us to underestimate the continuum level more at higher redshifts. 
Mock spectra, with continua blindly estimated following the same procedure as with the actual data, are used to quantify the mean fractional difference between the true and estimated continua, $\langle \Delta C/C_{\rm true} \rangle$ (where $\Delta C = C_{\rm est}-C_{\rm true}$), as function of redshift.
We find that, for hydrogen background photoionization rates \Gbkg~taken from the literature, $\Delta C/C_{\rm true}\approx1.58\times10^{-5}(1+z)^{5.63}$ over the redshift range $2\leq z \leq4.5$.
This continuum correction, however, itself is subject to some systematic bias -- since it depends on assumptions (particularly regarding \Gbkg) used to generate the mock spectra - which is unaccounted for in our statistical error bars.

\subsection{Metal Absorption}
Metals also contribute to the total absorption in the \Lya~forest.
In general, $\tau_{\rm eff} = \tau_{{\rm eff},\alpha} + \sum_{m} \tau_{{\rm eff},m}$, where $\tau_{{\rm eff},\alpha}$ arises purely from \Lya~absorption and $\tau_{{\rm eff},m}$ is due to metal $m$.
Based on the published results of \cite{2003ApJ...596..768S} and \cite{2005MNRAS.360.1373K}, we find that metals make up $(13-21\%,~6-9\%,~2-5\%)$ of $\tau_{\rm eff}$ at $z=(2,~3,~4)$.

\section{RESULTS}
\begin{figure}
\begin{center}
  \includegraphics[height=.36\textheight]{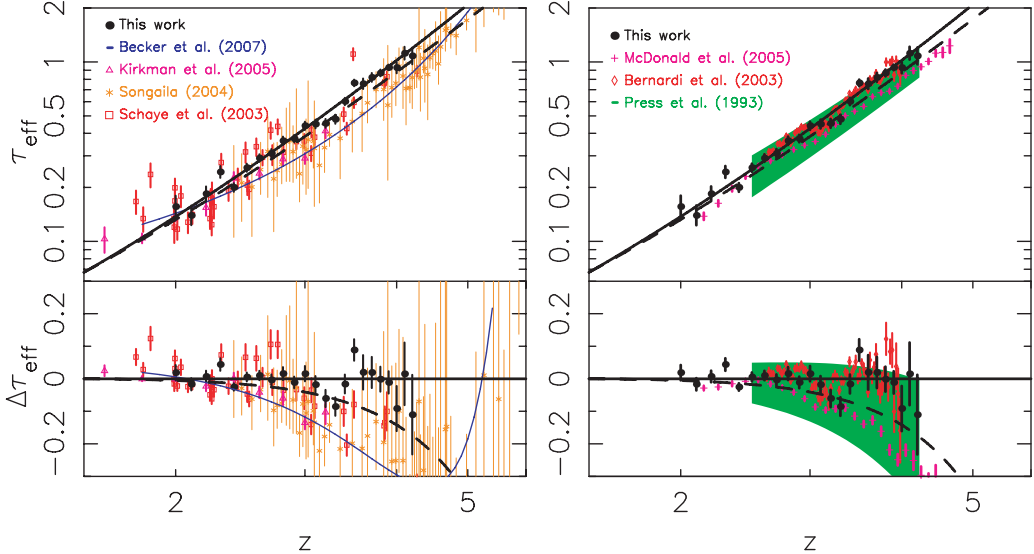}
  \caption{Comparison of our $\tau_{\rm eff}$ measurements with published results.
\emph{Left:} measurements in which the quasar continuum levels were estimated directly from high-resolution, high-signal-to-noise spectra, as in this work \citep{2003ApJ...596..768S, 2004AJ....127.2598S, 2005MNRAS.360.1373K, 2007ApJ...662...72B}.
\emph{Right:} estimates based extrapolating the quasar continuum from redward of \Lya~emission in low-resolution spectra \citep{1993ApJ...418..585P, 2003AJ....125...32B, 2005ApJ...635..761M}.
The solid black curves show the best-fit power law to our measurement corrected for continuum bias, but including metal absorption.
The dashed black curves indicate the best-fit power law to our measurement prior to correction for continuum bias.
Our measurement corrected for continuum bias is seen to trace the SDSS statistical measurement of \cite{2003AJ....125...32B}, who have first claimed detection of a feature at $z=3.2$, remarkably well.
This comparison is made quantitative in Figure \ref{compare us bernardi}.}
\label{measurements all}
\end{center}
\end{figure}
In Figure \ref{measurements all}, we compare our $\tau_{\rm eff}$ measurements with previously published measurements.
Our measurement corrected for continuum bias is seen to trace the SDSS statistical measurement of \cite{2003AJ....125...32B}, who have first claimed detection of a feature at $z=3.2$, remarkably well.
\begin{figure}
\begin{center}
  \includegraphics[height=.36\textheight]{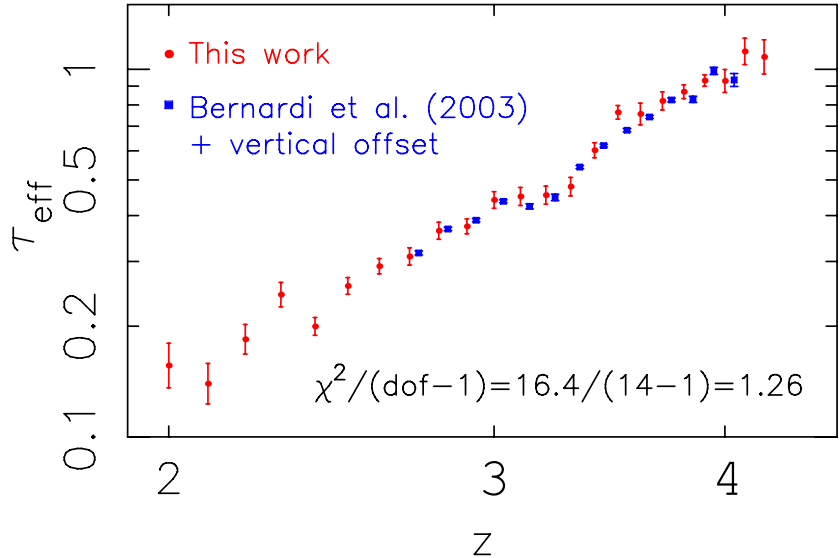}
  \caption{Comparison of our measurement of $\tau_{\rm eff}$ corrected for continuum bias (red circles) and of the \cite{2003AJ....125...32B} measurement (blue squares).
The \cite{2003AJ....125...32B} points were binned in the same $\Delta z=0.1$ redshift bins as our measurement.
The blue data points are centered on the same redshifts as the red ones, but have been slightly offset to the right for graphical clarity.
We have added a constant vertical offset in logarithmic space to the \cite{2003AJ....125...32B} points that minimizes the $\chi^{2}$ between the two data sets ($\Delta \log{\tau_{\rm eff}}=0.011$), since their statistical measurement may not be accurately normalized.
The resulting $\chi^{2}/dof=16.4/(14-1)=1.26$ indicates that the two data sets agree very well, with a $p$-value of 23\%.}
\label{compare us bernardi}
\end{center}
\end{figure}

\acknowledgments
CAFG is supported by a NSERC Postgraduate Fellowship.
JXP is supported by an NSF CAREER grant (AST-0548180) and by 
NSF grant AST-0709235.
This work was further supported in part by NSF grants ACI
96-19019, AST 00-71019, AST 02-06299, AST 03-07690, and AST 05-06556, and NASA ATP grants NAG5-12140, NAG5-13292, NAG5-13381, and NNG-05GJ40G.
Additional support was provided by the David and Lucile Packard, the Alfred P. Sloan, and the John D. and Catherine T. MacArthur Foundations.

\bibliographystyle{varenna}
\bibliography{references}

\end{document}